\documentclass[prb,twocolumn,superscriptaddress,showpacs,floatfix]{revtex4}

\usepackage{epsfig,amsmath,amssymb,latexsym}

\usepackage{subfigure}

\begin{document}

\title{Stabilizing the spiral order with spin-orbit coupling in an anisotropic triangular antiferromagnet}
\preprint{1}

\author{Xiao-Yong Feng}
\author{Xin Dong}
  \affiliation{Condensed Matter Group,
  Department of Physics, Hangzhou Normal University, Hangzhou 310036, China}

\author{Jianhui Dai}
  \affiliation{Condensed Matter Group,
  Department of Physics, Hangzhou Normal University, Hangzhou 310036, China}
  \affiliation{Department of Physics,
  Zhejiang University, Hangzhou 310027, China}


\begin{abstract}
We study the effects of spin-orbit coupling (SOC) on the large-U
Hubbard model on anisotropic triangular lattice at half-filling
using the Schwinger-boson method. We find that the SOC will in
general lead to a zero temperature condensation of the Schwinger
bosons with a single condensation momentum. As a consequence, the
spin-spin correlation vanishes along the $z$-axis but develops in
the $x$-$y$ plane, with the ordering vector being dramatically
dependent on the SOC. Moreover, the phase boundary of the magnetic
ordered state extends to the region of large spatial anisotropy with
increasing condensation density, demonstrating that the spiral order
is always stabilized by the SOC.
\end{abstract}

 \maketitle

Antiferromagnets on the triangular lattice represent a prototype
correlated systems where certain novel mangetic phases such as the
spin liquid state may emerge due to the quantum fluctuations and
geometric frustrations\cite{Anderson}. Some materials candidates of
the spin liquid, such as the organic
$\kappa-(BEDT-TTF)_2Cu_2(CN)_3$\cite{SL1} and
$EtMe_3Sb[Pd(dmit)_2]_2$\cite{SL2}, are the triangular-lattice
antiferromagnets, where the interchain coupling $J'$ is close to the
intrachain coupling $J$. Another layered quantum magnet,
$Cs_2CuCl_4$\cite{parameter}, has a relatively large spatial
anisotropy $\alpha\equiv J'/J\sim 0.34$. The inelastic neutron
scattering measurements reveal the possible spin liquid phase for
temperature above $T_N=0.62K$, while for $T<T_N$, the spiral order
is set up\cite{exp}. The emergent spiral order is attributed to the
interlayer coupling and the small Dzyaloshinskii-Moriya (DM)
interaction\cite{DM,DM3,DM1,DM2} which is put by hand in the
Heisenberg model. On the other hand, the SOC is ubiquitous in
materials whose crystal structures lack the inversion symmetry, and
$Cs_2CuCl_4$ just falls into this category.

Theoretically, though the SOC has been extensively studied for the
electronic systems with relatively small Coulomb interaction $U$, it
remains challenging to understand the effect of SOC when $U$ is
moderate or strong. In this paper, we study the Hubbard model on
anisotropic triangular lattice with the finite SOC and the infinite
$U$. The spatial anisotropy $\alpha$ in the studied model is an
important tuning parameter which interpolates the decoupled chains
($\alpha\rightarrow 0$) and the square lattice ($\alpha\rightarrow
\infty$). In the absence of the SOC,  a quasi one-dimensional spin
liquid phase and a two-dimensional magnetic phase emerge at the
half-filling in the two limiting cases, respectively. Thus by
increasing $\alpha$, it is natural to expect a critical $\alpha_c$,
separating the spin liquid and magnetic ordered states. Various
numerical calculations for finite clusters \cite{DMRG,VWF,FRG}
obtain $\alpha_c \sim 0.7-0.9 $, while the linear spin-wave
theory\cite{LSW} predicts a much smaller value $\alpha_c=0.27$. Here
we shall mainly focus on the magnetic ordered phase and study the
influence of the SOC by using the Schwinger-boson mean-field theory.

\begin{figure}[h]
\centering \includegraphics [width=6cm]{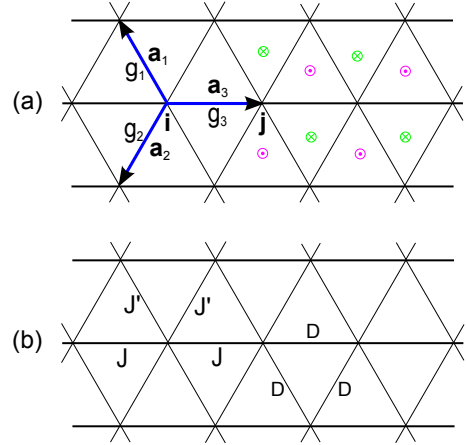} \caption{The
triangular lattice. (a)The bond vectors are
$\mathbf{a}_1$,$\mathbf{a}_2$ and $\mathbf{a}_3$. The electrons with
up spin feel the staggered flux as shown in the figure and the
electrons with down spin feel the flux with opposite direction.
(b)The spin coupling strength and the phase factor at each bond.}
\label{fig:triangular}
\end{figure}
Our starting point Hamiltonian is the Hubbard model on triangular
lattice with the SOC, defined by
\begin{eqnarray} H=-\sum_{\langle
\mathbf{i}\mathbf{j}\rangle,\sigma}[(t_{\mathbf{i}\mathbf{j}}-i\sigma
g_{\mathbf{i}\mathbf{j}})c_{\mathbf{i}\sigma}^{\dag}c_{\mathbf{j}\sigma}+h.c.]
+ U\sum_{\mathbf{i}}n_{\mathbf{i}\uparrow}n_{\mathbf{i}\downarrow}.
\end{eqnarray}
Where, $t_{\mathbf{i}\mathbf{j}}$ is the hopping energy between the
nearest neighboring bond $\langle \mathbf{i}\mathbf{j}\rangle$,
$c_{\mathbf{i}\sigma}$ annihilates an electron at $\mathbf{i}$ with
the spin index $\sigma=+/-$ for up/down spins,
$g_{\mathbf{i}\mathbf{j}}$ represents the strength of the SOC. We
shall at first drive an equivalent spin model with finite SOC for
large U and at half-filling. This has been recently realized for the
same model defined on the kagome lattice\cite{Mei}. Here the similar
approach is illustrated by the two-site problem. Noticed that the
bond $\langle \mathbf{i}\mathbf{j}\rangle$ is oriented, the SOC
brings an stagger flux through each triangle plaquette as shown in
Fig.\ref{fig:triangular}(a). Thus the first term can be re-expressed
by $-\tilde{t}_{12}\sum_{\sigma}(e^{-i\sigma
D_{12}}c_{1\sigma}^{\dag}c_{2\sigma}+h.c.)$, with
$\tilde{t}_{12}=\sqrt{t_{12}^2+g_{12}^2}$ and
$D_{12}=\arctan{(g_{12}/t_{12})}$. By performing the site and
spin-dependent $U(1)$ transformation:
$\tilde{c}_{1\sigma}=e^{i\sigma (D_{12}/2)}c_{1\sigma}$,
$\tilde{c}_{2\sigma}=e^{-i\sigma (D_{12}/2)}c_{2\sigma}$, the
corresponding Hamiltonian (at the large U limit) is mapped to the
spin Hamiltonian $J_{12}\bf{\tilde{S}_1}\cdot \bf{\tilde{S}_2}$ with
$J_{12}=\frac{4\tilde{t}_{12}^2}{U}$ and
$\mathbf{\tilde{S}_{i}}=\frac{1}{2}\sum_{\sigma\sigma'}
\tilde{c}^{\dag}_{i\sigma}\vec{\sigma}_{\sigma\sigma'}
\tilde{c}_{i\sigma'}$.

Noticed that the triangular lattice is not a bipartite lattice,
there exists no compatible $U(1)$ transformation for every site. For
a lattice model, we should transform back to the original fermion
operators. Therefore, the spin model for the strong-coupled Hubbard
model on the half-filled triangular lattice turns out to be
\begin{eqnarray}\label{spin}
H = \sum_{\langle
\mathbf{i}\mathbf{j}\rangle}J_{\mathbf{i}\mathbf{j}}
\left[\frac{1}{2}(e^{-2iD_{\mathbf{i}\mathbf{j}}}
S_{\mathbf{i}}^{+}S_{\mathbf{j}}^{-}+
e^{2iD_{\mathbf{i}\mathbf{j}}}S_{\mathbf{i}}^{-}
S_{\mathbf{j}}^{+})+S^{z}_{\mathbf{i}}S^{z}_{\mathbf{j}} \right],
\end{eqnarray}
where $S^{\pm}=S^x\pm iS^y$. The SU(2) symmetry of  the above spin
model is broken by the SOC. And its antisymmetric (the DM-term ) and
symmetric parts appear with explicit SOC-dependent coupling
strengths.

We then study the Hamiltonian (\ref{spin}) using the Schwinger-boson
mean-field theory\cite{Assa}. The spin operators are represented by
boson operators $S^{+}=b^{\dag}_{\uparrow}b_{\downarrow} $,
$S^{-}=b^{\dag}_{\downarrow}b_{\uparrow}$,
$S^{z}=\frac{1}{2}(b^{\dag}_{\uparrow}b_{\uparrow}-b^{\dag}_{\downarrow}b_{\downarrow})$,
with the constraint
$b^{\dag}_{\uparrow}b_{\uparrow}+b^{\dag}_{\downarrow}b_{\downarrow}=1$.
We introduce the mean-fields
$
  \psi_{n\sigma}=-i\sigma\langle b_{\mathbf{i}\sigma}
b_{\mathbf{i}+\mathbf{a}_n\bar{\sigma}}\rangle $, with
$\mathbf{a}_1=a_0(-\frac{1}{2},\frac{\sqrt{3}}{2})$,
$\mathbf{a}_2=a_0(-\frac{1}{2},-\frac{\sqrt{3}}{2})$,
$\mathbf{a}_3=-(\mathbf{a}_1+\mathbf{a}_2)=a_0(1,0)$, and $a_0$ is
the lattice constant, as marked in Fig.\ref{fig:triangular}(a). The
obtained mean-field Hamiltonian is
\begin{eqnarray}
H &=&\frac{i}{2}\sum_{\mathbf{i}n\sigma}\bar{\sigma}J_{n}(\psi_{n\sigma}+e^{2i\bar{\sigma}D_n}\psi_{n\bar{\sigma}})b^{\dag}_{\mathbf{i}\sigma}b^{\dag}_{\mathbf{i}+\mathbf{a}_n\bar{\sigma}}+h.c.\nonumber\\
&+&\lambda\sum_{\mathbf{i}}(\sum_\sigma
b^{\dag}_{\mathbf{i}\sigma}b_{\mathbf{i}\sigma}-1),
\end{eqnarray}
where the $\lambda$-term is introduced to impose the constraint.


In the momentum space, the Hamiltonian has the form
\begin{eqnarray}
H=\sum_\mathbf{k}(b^{\dag}_{\mathbf{k}\uparrow},
b_{-\mathbf{k}\downarrow})\left(
                                                  \begin{array}{cc}
                                                    \lambda & A_{\mathbf{k}} \\
                                                    A_{\mathbf{k}}^{*} & \lambda \\
                                                  \end{array}
                                                \right)\left(
                                                         \begin{array}{c}
                                                           b_{\mathbf{k}\uparrow} \\
                                                           b_{-\mathbf{k}\downarrow}^{\dag} \\
                                                         \end{array}
                                                       \right)-2N\lambda,
\end{eqnarray}
where $N=N_1\times N_2$ is the site number of the lattice (
$\mathbf{a}_1$ and $\mathbf{a}_2$ are chosen as the two primitive
translation vectors), $
A_{\mathbf{k}}=\frac{i}{2}\sum_{n\sigma}\bar{\sigma}
J_n(\psi_{n\sigma}+e^{2i\bar{\sigma}D_n}\psi_{n\bar{\sigma}})e^{i\sigma\mathbf{k}\cdot\mathbf{a}_n}$.
We assume that $A_{\mathbf{k}}$ is a real number, because the phase
factor of $A_{\mathbf{k}}$ can be gauged away by the U(1)
transformation. Due to the time-reversal symmetry, we also have
$|\psi_{n\uparrow}|=|\psi_{n\downarrow}|$. Thus we use the ansatz
$\psi_{n\sigma}=\psi_n e^{i\sigma\vartheta_n}$, and obtain $
A_{\mathbf{k}}=2\sum_{n}J_n\sin{(\mathbf{k}\cdot\mathbf{a}_n-D_n)\cos{(D_n+\vartheta_n)}}\psi_n
$.

The mean-field Hamiltonian  Eq.(4) is diagonalized by the bosonic
Bogliulov transformation,
$b_{\mathbf{k}\uparrow}=u_\mathbf{k}\alpha_{\mathbf{k}\uparrow}
  +v_\mathbf{k}\alpha_{\mathbf{k}\downarrow}^{\dag}$,
$b_{-\mathbf{k}\downarrow}=u_\mathbf{k}\alpha_{\mathbf{k}\downarrow}
+v_\mathbf{k}\alpha_{\mathbf{k}\uparrow}^{\dag}$, with
$u_\mathbf{k}^2=\frac{\lambda}{2\omega_{\mathbf{k}}}+\frac{1}{2}$,
$v_\mathbf{k}^2=\frac{\lambda}{2\omega_{\mathbf{k}}}-\frac{1}{2}$,
and
$u_\mathbf{k}v_\mathbf{k}=-\frac{A_\mathbf{k}}{2\omega_\mathbf{k}}$.
Then
\begin{eqnarray}
H=\sum_{\mathbf{k}\sigma}\omega_{\mathbf{k}}\alpha_{\mathbf{k}\sigma}^{\dag}\alpha_{\mathbf{k}\sigma}
+\sum_{\mathbf{k}}(\omega_{\mathbf{k}}-2\lambda),
\end{eqnarray}
where the quasiparticle dispersion
$\omega_{\mathbf{k}}=\sqrt{\lambda^2-A_{\mathbf{k}}^2}$. A stable
ground state requires $\lambda\geq|A_{\mathbf{k}}|$. The free energy
is given by
\begin{eqnarray}
F =
\frac{2}{\beta}\sum_{\mathbf{k}}\ln{(1-e^{-\beta\omega_\mathbf{k}})}+\sum_{\mathbf{k}}(\omega_{\mathbf{k}}-2\lambda)
\end{eqnarray}
with $\beta=\frac{1}{k_BT}$. The Lagrangian multiplier $\lambda$ is
determined by optimizing the free energy, leading to
\begin{eqnarray}\label{constraint}
\frac{1}{N}\sum_{\mathbf{k}}\frac{\lambda}{\omega_\mathbf{k}}\left(n_b(\omega_\mathbf{k})+\frac{1}{2}\right)=1,
\end{eqnarray}
where
$n_b(\omega_\mathbf{k})=\frac{1}{e^{\beta\omega_\mathbf{k}}-1}$ is
the Bose distribution function. The mean-fields $\psi_n$ and
$\vartheta_n$ are then calculated self-consistently through
following equations,
\begin{eqnarray}\label{equations}
\frac{i}{N}\sum_{\mathbf{k}}\frac{A_\mathbf{k}}{\omega_\mathbf{k}}\left(n_b(\omega_\mathbf{k})+\frac{1}{2}\right)
e^{-i\mathbf{k}\cdot\mathbf{a}_n}=\psi_ne^{i\vartheta_n}.
\end{eqnarray}

Keeping in mind that $J_n = (t_n^2+g_n^2)/U$ and $D_n =
\arctan{(g_n/t_n)}$,  we set $J_1=J_2=J'$, $J_3=J$, and
$D_1=D_2=D_3=D$ as illustrated in Fig.\ref{fig:triangular}(b). Thus
the spatial anisotropy is measured by $\alpha\equiv J'/J$. Three
special cases in the parameter space are $(1)$ $\alpha=0$, the
decoupled chain's limit; $(2)$ $\alpha=\infty$, the square lattice
limit; and $(3)$ $\alpha=1$, the isotropic point. Notice that with
this choice of parameters, the mean-field equations are invariant
under the exchange of the bond index $1$ and $2$. Therefore, we have
$\psi_1=\psi_2$ and $\vartheta_1=\vartheta_2$.

According to the analysis with functional integral method\cite{AA},
no bosons can condense in the two-dimensional lattices at finite
temperature. This is consistent with the Mermin-Wagner
theorem\cite{Mermin}. 
In the case of a gapped energy spectrum,
i.e. $\min{\omega(\mathbf{k})}\neq0$, there is no Bose-Einstein
condensation and the ground state is a spin liquid. Otherwise, if
the spectrum is gapless, the bosons can condense on the lowest
energy state, implying a long-range magnetic order.\cite{Shen} The
relation between magnetic order and Bose-Einstein condensation can
be drawn from the spin-spin correlation whose
diagonal components are given by
\begin{eqnarray}
&&\langle
S^x_\mathbf{0}S^x_\mathbf{i}\rangle=\langle S^y_\mathbf{0}S^y_\mathbf{i}\rangle
=\frac{1}{2}Re(f_\mathbf{i}^2+g_\mathbf{i}^2), \\
&&\langle
S^z_\mathbf{0}S^z_\mathbf{i}\rangle=\frac{1}{2}(|f_\mathbf{i}|^2-|g_\mathbf{i}|^2),
\end{eqnarray}
where,
\begin{eqnarray}
f_\mathbf{i}&=&\frac{1}{N}\sum_{\mathbf{k}}\frac{\lambda}{\omega_\mathbf{k}}\left(n_b(\omega_\mathbf{k})+\frac{1}{2}\right)e^{i\mathbf{k}\cdot\mathbf{i}},\\
g_\mathbf{i}&=&\frac{1}{N}\sum_{\mathbf{k}}\frac{A_\mathbf{k}}{\omega_\mathbf{k}}\left(n_b(\omega_\mathbf{k})+\frac{1}{2}\right)e^{i\mathbf{k}\cdot\mathbf{i}}.
\end{eqnarray}

It is important to recall that when $D=0$, i.e., in the absence of
the SOC, $A_{-\mathbf{k}}= -A_\mathbf{k}$, the SU(2) symmetry is
restored. Then there always exists a pair of zero modes of
$\omega(\mathbf{k})$ with $\mathbf{k}=\pm\mathbf{k}^*$, condensed at
zero temperature. However, for non-zero SOC, $D\neq 0$, we find that
$A_{-\mathbf{k}}\neq -A_\mathbf{k}$ in general and
$\omega(\mathbf{k})$ has only a single zero mode $\mathbf{k}^*$ in
the first Brillouin zone, so the expression for $x-$ or
$y-$component of the spin-spin correlation is different from that of
the z-component.  
According to Eq. (10), the spin-spin correlation between the
$z$-components vanishes at zero temperature when the distance
between two spins is sufficiently large. Moreover, there exits one
independent nonvanishing off-diagonal component, $ \langle
S^y_\mathbf{0}S^x_\mathbf{i}\rangle=-\langle
S^x_\mathbf{0}S^y_\mathbf{i}\rangle=\frac{1}{2}Im(f_\mathbf{i}^2+g_\mathbf{i}^2)
$.   All these features are in contrast with the cases studied
previously\cite{Sanker,Shen,AA}.

Now we discuss the numerical results for the infinite system. By
converting the sum in the mean-field equations (\ref{constraint})
and (\ref{equations}) into integrals and denoting the contribution
from the Bose condensate as $b_0$, the mean-field equations for
numerical performance are
\begin{eqnarray}
\int{\frac{d^2\mathbf{k}}{(2\pi)^2}\frac{\lambda
}{2\omega_\mathbf{k}}}=1-b_0, \\
i\int{\frac{d^2\mathbf{k}}{(2\pi)^2}}
\frac{A_\mathbf{k}e^{-i\mathbf{k}\cdot\mathbf{a}_n}}
{2\omega_\mathbf{k}}+iB_0=\psi_ne^{i\vartheta_n},
\end{eqnarray}
where
$B_0=\frac{1}{\lambda}A_\mathbf{k^*}e^{-i\mathbf{k}^*\cdot\mathbf{a}_n}b_0$
if $b_0>0$, and $B_0=0$ if $b_0<0$. To have a gapless spectrum,
$\lambda$ is always fixed at the largest value of $|A_\mathbf{k}|$.
If the solution is associated with a negative $b_0$, we can always
fulfill the constraint (\ref{constraint}) by tuning up $\lambda$
slightly. In this case, the energy spectrum is gapped and the system
is in the spin liquid phase. If there is a solution with a positive
$b_0$, then the system is in the condensation phase. The magnetic
order can be obtained from the spin-spin correlation functions. They
are $\langle S^x_\mathbf{0}S^x_\mathbf{i}\rangle=\langle
S^y_\mathbf{0}S^y_\mathbf{i}\rangle=b_0^2\cos{(2\mathbf{k}^*\cdot\mathbf{i})}$
and $\langle S^y_\mathbf{0}S^x_\mathbf{i}\rangle=-\langle
S^x_\mathbf{0}S^y_\mathbf{i}\rangle=b_0^2\sin{(2\mathbf{k}^*\cdot\mathbf{i})}$.
By denoting $\mathbf{k}^*=(k_1^*,k_2^*)$, we have the ordering wave
vectors, along the directions of $\mathbf{a}_1$, $\mathbf{a}_2$ and
$\mathbf{a}_3$, being $2k_1^*$, $2k_2^*$
and $-2(k_1^*+k_2^*)$, modulo-divided by $2\pi$, respectively.

\begin{figure}[h]
\centering \includegraphics [width=9cm]{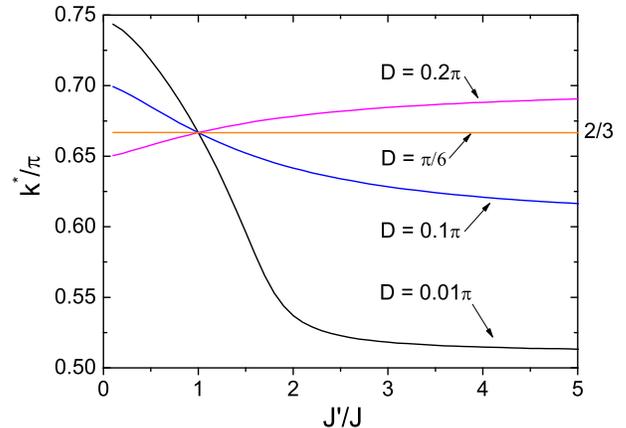} \caption{The
condensation momentum $\mathbf{k}^*=(k^*, k^*)$ as a function of
$J'/J$ for $D=0.01\pi$, $0.1\pi$, $\frac{1}{6}\pi$ and $0.2\pi$
respectively.} \label{fig:k_J}
\end{figure}

The numerical solutions for $\mathbf{k}^*$ show that
$k_1^*=k_2^*=k^*$. The curves of $k^*$ as a function of
$\alpha\equiv J'/J$ are plotted in Fig.\ref{fig:k_J}. When $D$ is
very small, the ordering wave vector along the chain tends to $\pi$
near the decoupled-chain limit ($\alpha\ll1$), indicating an
antiferromagnetic order. While when $\alpha\gg1$, the ordering wave
vectors along the directions of $\mathbf{a}_1$, $\mathbf{a}_2$ tend
to $\pi$, $\pi$, respectively, reproducing the checkerboard or Neel
ordering for the unfrustrated antiferromagnet in a square lattice.
Between the two limits the spiral order develops and the order
parameter oscillates with the distance. With the increase of $D$,
the curves of $k^*$ change dramatically. Especially, when
$D=\frac{1}{6}\pi$, $k^*=\frac{2}{3}\pi$ is independent of $\alpha$.
For generic $D$, the limiting values of $k^*$ with small and large
$\alpha$ are $\frac{3}{4}\pi-\frac{1}{2}D$ and $\frac{1}{2}\pi+D$
respectively. It is also interesting to note that at the isotropic
point ($\alpha=1$), $k^*$ always equals to $\frac{2}{3}\pi$
regardless of $D$.

\begin{figure}[h]
\centering \includegraphics [width=9cm]{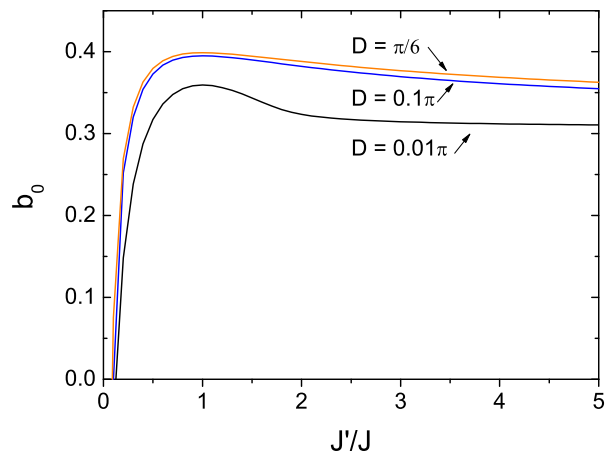} \caption{The Bose
condensation density $b_0$ as a function of $J'/J$ for $D=0.01\pi$,
$0.1\pi$ and $\frac{1}{6}\pi$ respectively.} \label{fig:d_J}
\end{figure}

The square-root of the amplitude of the spin-spin correlation
function, characterizing the strength of the magnetic order, is
exactly the Bose condensation density $b_0$. In the case of the
antiferromagnetic order, $b_0$ is equal to the sublattice
magnetization. It is also a function of $D$ with period $\pi/3$
since the SOC can be related to the phase factor in the triangular
lattice. As shown in Fig \ref{fig:d_J}, $b_0$ is enhanced
monotonically when $D$ increases up to $\pi/6$. For each $D$, $b_0$
takes a maximum value at $\alpha=1$.

\begin{figure}[h]
\centering \includegraphics [width=9cm]{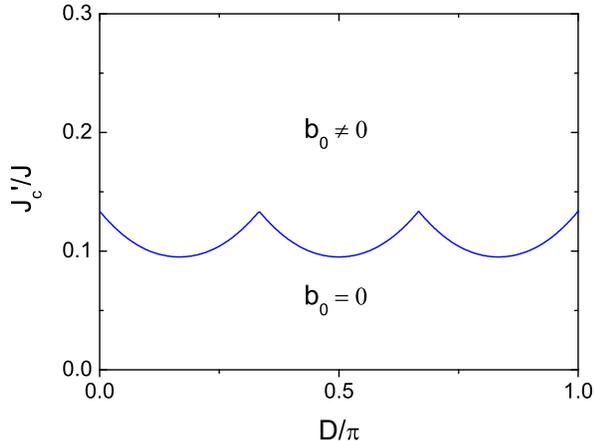} \caption{The
mean-field phase boundary between the magnetic ordered state  and
the spin liquid state.} \label{fig:Jc_D}
\end{figure}

For each fixed $D$ there is a critical value $\alpha_c\equiv J'_c/J$
separating the spin liquid phase from the magnetic ordered phase.
For $\alpha<\alpha_c$, $b_0=0$ and all the spin-spin correlation
functions decay exponentially to zero as the distance
$\mathbf{i}\rightarrow\infty$. For $\alpha>\alpha_c$, the bosons
condense on the state with momentum $\mathbf{k}^*$. Fig.
\ref{fig:Jc_D} plots the critical line of the quantum phase
transition. As a function of $D$, the critical line oscillates with
period $\pi/3$ as expected. In realistic materials, the strength of
the SOC is usually small compared to the hopping integral and the
large $D$ in the phase diagram is unphysical. However, large value
of D is achievable in the optical flux lattice\cite{Op}. We find
that the nonzero $D$ (within the half-period $\pi/6$) pushes the
boundary to the region with a smaller $\alpha$, thus it effectively
enhances the magnetic ordering. This feature is compatible with the
$D$-dependence of $b_0$. When $D=\frac{\pi}{6}$, $\alpha_c$ reaches
the minimum value $0.095$. This result implies that with relatively
large SOC the spiral magnetic order can emerge in the triangular
lattice which has a tendency of reducing dimensionality as featured
by small $\alpha$.

Finally, we remark that our mean-field treatment emphasizes the
magnetic ordering but is not adequate to capture the disordered
state which may appear in the region with the moderate spatial
anisotropy.  Focusing on the spiral ordered phase with small
anisotropy, it is possible that the quantum fluctuations could blur
the difference between the solutions with two condensation momenta
and a single momentum, giving rise to a finite spin-spin correlation
between $z$-components. While this issue can be clarified in the
future by taking into account the quantum fluctuations above the
mean-field solution, it is robust that the SOC induces a nonzero
spin correlation between the $x$- and $y$-components, suppresses the
spin correlation between the $z$-components and favors the
establishment of the spiral order.

In summary, a large-U Hubbard model with the SOC on an spatial
anisotropic lattice at half-filling is studied using the Schwinger-
boson method. With the participation of the SOC, the condensation
momentum of the Schwinger bosons have only a single zero mode which
in turn leads to a finite spin-spin correlation between $x$- and
$y$-components and a vanishing spin-spin correlation between
$z$-components at large distance. The SOC also shifts the phase
boundary between the magnetic ordered state and the spin liquid
state to the larger anisotropy side, and stabilizes the spiral
magnetic order. Our results provide an alternative understanding of
the spiral order observed in some materials like $Cs_2CuCl_4$ where
the triangular lattice has a relatively small $\alpha$.

This work was supported in part by the NSFC, the NSF of Zhejiang
Province, and the 973 Project of the MOST.


\end{document}